\documentclass[conference]{IEEEtran}

\usepackage{graphicx}
\usepackage{amsmath}
\begin{document}
%
\title{Energy Efficiency of Network Cooperation for Cellular Uplink Transmissions}
%
%
%

\author{\IEEEauthorblockN{Yulong Zou, Jia Zhu, and Baoyu Zheng}
\IEEEauthorblockA{Inst. of Signal Process. and Transm., Nanjing Univ. Post \& Telecomm., Nanjing, P. R. China\\
Email: \{Yulong.Zou, Jiazhu2010\}@gmail.com, zby@njupt.edu.cn}

}

\maketitle

\begin{abstract}
There is a growing interest in energy efficient or so-called ``green'' wireless communication to reduce the energy consumption in cellular networks. Since today's wireless terminals are typically equipped with multiple network access interfaces such as Bluetooth, Wi-Fi, and cellular networks, this paper investigates user terminals cooperating with each other in transmitting their data packets to a base station (BS) by exploiting the multiple network access interfaces, referred to as \emph{inter-network cooperation}, to improve the energy efficiency in cellular uplink transmission. Given target outage probability and data rate requirements, we develop a closed-form expression of energy efficiency in Bits-per-Joule for the inter-network cooperation by taking into account the path loss, fading, and thermal noise effects. Numerical results show that when the cooperating users move towards to each other, the proposed inter-network cooperation significantly improves the energy efficiency as compared with the traditional non-cooperation and intra-network cooperation. This implies that given a certain amount of bits to be transmitted, the inter-network cooperation requires less energy than the traditional non-cooperation and intra-network cooperation, showing the energy saving benefit of inter-network cooperation.

\end{abstract}

\begin{IEEEkeywords}
Energy efficiency, network cooperation, cellular network, outage probability, green communication.
\end{IEEEkeywords}

\IEEEpeerreviewmaketitle

\section{Introduction}
%
%
%
%
\IEEEPARstart In wireless communication, path loss and fading are two major issues to be addressed in order to improve the quality of service (QoS) of various applications (voice, data, multimedia, etc.) [1]. Typically, the channel path loss and fading are determined by many factors including the terrain environment (urban or rural), electromagnetic wave frequency, distance between the transmitter and receiver, antenna height, and so on. In the case of large path loss and deep fading, more transmit power is generally required to maintain a target QoS requirement. For example, in cellular networks, a user terminal at the edge of its associated cell significantly drains its battery energy much faster than that located at the cell center. Therefore, it is practically important to study energy-efficient or so-called green wireless communication to reduce the energy consumption in cellular networks, especially for cell-edge users [2], [3].

User cooperation has been recognized as an effective means to achieve spatial diversity and improve wireless transmission performance. In [4], the authors studied cooperative users in relaying each other's transmission to a common destination and examined the outage performance of various relaying protocols (i.e., fixed relaying, selective relaying, and incremental relaying). In [5], the Alamouti space-time coding was examined for regenerative relay networks, where the relay first decodes its received signals from a source node and then re-encodes and forwards its decoded signal to a destination node. In [6], the authors studied the space-time coding in amplify-and-forward relay networks and proposed a distributed linear dispersion code for the cooperative relay transmissions. More recently, user cooperation has been exploited in emerging cognitive radio networks with various cooperative relaying protocols for spectrum sensing and cognitive transmissions [7], [8]. It is worth noting that the aforementioned intra-network user cooperation typically operates with a single network access interface.

In today's wireless networks, a user terminal (e.g., smart phone) is typically equipped with multiple network access interfaces to support both short-range communication (via e.g. Bluetooth and Wi-Fi) and long-range communication (via e.g. cellular networks) [9], with different radio characteristics in terms of coverage area and energy consumption. Specifically, the short-range networks provide local-area coverage with low energy consumption, whereas cellular networks offer wider coverage with higher energy consumption. This implies that different radio access networks complement each other in terms of the network coverage and energy consumption. In order to take advantages of different existing radio access networks, it is of high practical interest to exploit the multiple network access interfaces assisted user cooperation, termed inter-network cooperation. In this paper, we study the inter-network cooperation to improve energy efficiency of the cellular uplink transmission with the assistance of a short-range communication network.

The main contributions of this paper are summarized as follows. First, we present an inter-network cooperation framework in a heterogeneous environment consisting of different radio access networks (e.g., a short-range communication network and a cellular network). Then, we examine distributed space-time coding techniques for the proposed scheme and conduct the energy efficiency analysis with target outage probability and data rate constraints by taking into account the path loss, fading, and thermal noise effects. For the purpose of comparison, we also examine the energy efficiency of two benchmark schemes, including the traditional non-cooperation and existing intra-network cooperation (i.e., the user terminals cooperate via a common cellular network interface) [5], and show the advantage of proposed inter-network cooperation in terms of energy saving.

The remainder of this paper is organized as follows. Section II describes the system model and presents the traditional non-cooperation, intra-network cooperation and inter-network cooperation schemes. Next, we conduct the energy efficiency analysis with target outage probability and data rate requirements in Section III, followed by Section IV where numerical energy efficiency results are provided. Finally, we make some concluding remarks in Section V.

\section{Cellular Uplink Transmission Based on Network Cooperation}
In this section, we first present the system model of a heterogeneous network environment, where user terminals are assumed to have multiple radio access interfaces including a short-range communication interface and a cellular access interface. Then, we propose the inter-network cooperation scheme by exploiting the short-range network to assist cellular uplink transmissions as well as two baseline schemes for comparison.
\subsection{System Model}
Fig. 1 shows the system model of a heterogeneous network consisting of a base station (BS) and two user terminals as denoted by U1 and U2, each equipped with a short-range communication interface and a cellular access interface. The two users are assumed to cooperate with each other in transmitting to the BS. Since U1 and U2 are equipped with a short-range communication interface (e.g., Bluetooth), they are able to establish a short-range cooperative network to assist their cellular uplink transmissions and improve the overall energy efficiency. To be specific, we first allow U1 and U2 to communicate with each other and exchange their uplink data packets through the short-range network. Once U1 and U2 obtain each other's data packets, they can employ distributed space-time coding to transmit their data packets to BS by sharing their antennas. Note that the proposed scheme uses two different networks (i.e., a short-range network and a cellular network), which is thus termed inter-network cooperation and differs from traditional user cooperation schemes [4]-[6] that operate in a homogeneous network environment with one single radio access interface. Under the cellular network setup, the traditional user cooperation requires a user terminal to transmit its signal over a cellular spectrum band to its partner that then forwards the received signal to BS. This comes at the cost of low cellular spectrum utilization efficiency, since two orthogonal channels are required to complete one packet transmission from a user terminal to BS via its partner. In contrast, the inter-network cooperation allows a user terminal to transmit its signal to its partner using a short-range network (e.g., Bluetooth) over an industrial, scientific and medical (ISM) band, instead of using a cellular band. This thus saves cellular spectrum resources and can significantly improve the cellular spectrum utilization as compared with the traditional intra-network cooperation.
\begin{figure}
  \centering
  {\includegraphics[scale=0.6]{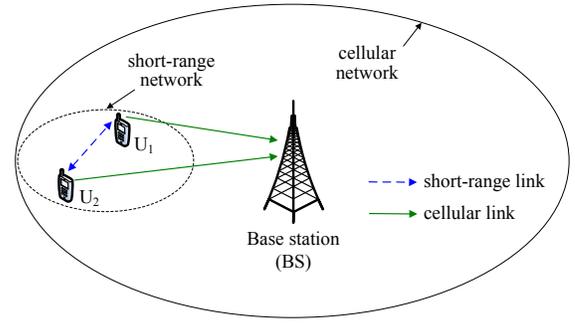}\\
  \caption{A cellular network consists of a base station (BS) and two user terminals that are equipped with multiple radio access interfaces (i.e., a short-range communication interface and a cellular access interface).}\label{Fig1}}
\end{figure}

In Fig. 1, a short-range network is considered to assist the cellular transmissions from U1 and U2 to BS. At present, there are two main short-range communication networks, i.e., Bluetooth and Wi-Fi, which typically operate at $2.4{\textrm{\emph{GHz}}}/5{\textrm{\emph{GHz}}}$ [10], [11]. However, cellular networks generally operate at $900{\textrm{\emph{MHz}}}/1800{\textrm{\emph{MHz}}}$. Thus, we consider a general channel model [1] that incorporates the radio frequency, path loss and fading effects in characterizing wireless transmissions in both types of networks as ${P_{Rx}} = {P_{Tx}}{\left(\dfrac{\lambda }{{4\pi d}}\right)^2}{G_{Tx}}{G_{Rx}}|h{|^2}$, where ${P_{Rx}}$ is the received power, ${P_{Tx}}$ is the transmitted power, $\lambda $ is the carrier wavelength, $d$ is the transmission distance, ${G_{Tx}}$ is the transmit antenna gain, ${G_{Rx}}$ is the receive antenna gain, and $h$ is the channel fading coefficient. Throughout this paper, we consider a Rayleigh fading model to characterize the channel fading, i.e., $|h{|^2}$ is modeled as an exponential random variable with mean $\sigma _h^2$. Also, all receivers are assumed with the circularly symmetric complex Gaussian (CSCG) distributed thermal noise with zero mean and noise variance $\sigma _n^2$. As reported in [1], the noise variance $\sigma _n^2$ is modeled as $\sigma _n^2 = \kappa TB$, where $\kappa $ is the Boltzmann constant, $T$ is the system temperature, and $B$ is the channel bandwidth. Denoting ${N_0} = \kappa T$ (called noise power spectral density) and considering room temperature $T = 290K$, we can easily obtain ${N_0} =  - 174{\textrm{\emph{dBm/Hz}}}$.

\subsection{Traditional Non-Cooperation}

\begin{figure}
  \centering
  {\includegraphics[scale=0.65]{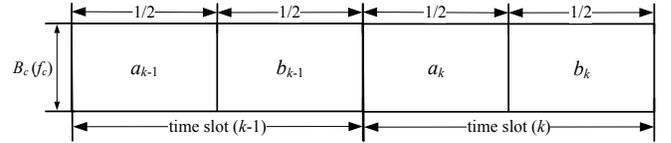}\\
  \caption{Time slot structure of the traditional non-cooperation for cellular uplink transmissions from U1 and U2 to BS employing time division multiple access (TDMA), where ${f_c}$ and ${B_c}$ are the cellular carrier frequency and spectrum bandwidth, respectively.}\label{Fig2}}
\end{figure}

First, consider the traditional non-cooperation as a baseline for comparison. Without loss of generality, let ${a_k}$ and ${b_k}$ denote transmit signals of U1 and U2, respectively, in time slot $k$. Fig. 2 shows the slot structure of the traditional scheme for cellular uplink transmissions based on time-division-multiple-access (TDMA), where ${f_c}$ and ${B_c}$ represent the cellular carrier frequency and spectrum bandwidth, respectively. As shown in Fig. 2, U1 and U2 take turns in accessing the cellular spectrum to transmit their signals ${a_k}$ and ${b_k}$ at data rates $R_1$ and $R_2$ in bits per second, respectively. Consider that U1 transmits ${a_k}$ with power ${P_1}$ and data rate ${R_1}$. Thus, the received signal-to-noise ratio (SNR) at BS from U1 is given by
\begin{equation}\label{equa1}
\gamma _{1b}^T = \dfrac{{{P_1}}}{{{N_0}{B_c}}}{\left(\dfrac{{{\lambda _c}}}{{4\pi {d_{1b}}}}\right)^2}{G_{U1}}{G_{BS}}|{h_{1b}}{|^2},
\end{equation}
where the superscript $T$ stands for `traditional', ${\lambda _c}=c/f_c$ is the cellular carrier wavelength, $c$ is the speed of light, ${B_c}$ is the cellular spectrum bandwidth, ${d_{1b}}$ is the transmission distance from U1 to BS, ${G_{U1}}$ is the transmit antenna gain at U1, ${G_{BS}}$ is the receive antenna gain at BS, ${h_{1b}}$ is the fading coefficient of the channel from U1 to BS. Similarly, considering that U2 transmits ${b_k}$ with power ${P_2}$ and data rate ${R_2}$, we can obtain the received SNR at BS from U2 as
\begin{equation}\label{equa2}
\gamma _{2b}^T = \dfrac{{{P_2}}}{{{N_0}{B_c}}}{\left(\dfrac{{{\lambda _c}}}{{4\pi {d_{2b}}}}\right)^2}{G_{U2}}{G_{BS}}|{h_{2b}}{|^2},
\end{equation}
where ${d_{2b}}$ is the transmission distance from U2 to BS, ${G_{U2}}$ is the transmit antenna gain at U2, and ${h_{2b}}$ is the fading coefficient of the channel from U2 to BS.
\subsection{Proposed Inter-Network Cooperation}
\begin{figure}
  \centering
  {\includegraphics[scale=0.65]{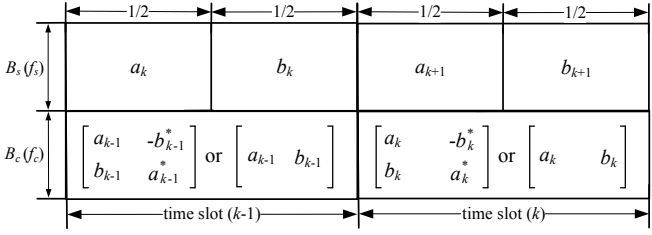}\\
  \caption{Time slot structure of the proposed inter-network cooperation scheme for cellular uplink transmissions from U1 and U2 to BS, where ${f_s}$ and ${B_s}$ are the carrier frequency and spectrum bandwidth of a short-range communication network, and ${f_c}$ and ${B_c}$ are the cellular carrier frequency and spectrum bandwidth, respectively.}\label{Fig3}}
\end{figure}

This subsection describes the proposed inter-network cooperation scheme for cellular uplink transmissions from U1 and U2 to BS. As shown in Fig. 3, we illustrate the time slot structure of the proposed scheme, where ${f_s}$ and ${B_s}$, respectively, represent the carrier frequency and spectrum bandwidth of a short-range communication network. The following details the cooperation process in transmitting ${a_k}$ and ${b_k}$ from U1 and U2 to BS. First, we allow U1 and U2 to exchange their signals in time slot $k-1$ over the short-range communication network, i.e., U1 transmits ${a_k}$ to U2 with power ${P_{1,s}}$ and data rate ${R_1}$ during the first half of slot $k-1$ and U2 transmits ${b_k}$ to U1 with power ${P_{2,s}}$ and data rate ${R_2}$ during the second half of slot $k-1$. Assuming that the gain of an antenna for the short-range communication is the same as that for the cellular transmission, the received SNR at U2 from U1 is given by
\begin{equation}\label{equa3}
\gamma _{12}^{NC} = \dfrac{{{P_{1,s}}}}{{{N_0}{B_s}}}{\left(\dfrac{{{\lambda _s}}}{{4\pi {d_{12}}}}\right)^2}{G_{U1}}{G_{U2}}|{h_{12}}{|^2},
\end{equation}
where the superscript $NC$ stands for `network cooperation', ${\lambda _s}=c/f_s$ is the carrier wavelength of the short-range communication, ${d_{12}}$ is the transmission distance from U1 to U2, and ${h_{12}}$ is the fading coefficient of the channel from U1 to U2. Also, the received SNR at U1 from U2 can be written as
\begin{equation}\label{equa4}
\gamma _{21}^{NC} = \dfrac{{{P_{2,s}}}}{{{N_0}{B_s}}}{\left(\dfrac{{{\lambda _s}}}{{4\pi {d_{21}}}}\right)^2}{G_{U1}}{G_{U2}}|{h_{21}}{|^2},
\end{equation}
where ${d_{21}}$ is the transmission distance from U2 to U1, and ${h_{21}}$ is the fading coefficient of the channel from U2 to U1. For notational convenience, let $\theta  = 1$ denote the case that both U1 and U2 succeed in decoding each other's signals and $\theta  = 2$ denote the other case that either U1 or U2 (or both) fails to decode. In the case of $\theta  = 1$, we adopt Alamouti space-time coding [12] for U1 and U2 in transmitting ${a_k}$ and ${b_k}$ to BS in time slot $k$ over a cellular spectrum band, where the transmit power values of U1 and U2 are denoted by ${P_{1,c}}$ and ${P_{2,c}}$, respectively. Specifically, during the first half of time slot $k$, U1 and U2, respectively, transmit ${a_k}$ and ${b_k}$ simultaneously, i.e., the received signal at BS is given by
\begin{equation}\label{equa5}
\begin{split}
{y_1} =& \sqrt {{P_{1,c}}{(\dfrac{{{\lambda _c}}}{{4\pi {d_{1b}}}})^2}{G_{U1}}{G_{BS}}} {h_{1b}}{a_k} \\
&+ \sqrt {{P_{2,c}}{(\dfrac{{{\lambda _c}}}{{4\pi {d_{2b}}}})^2}{G_{U2}}{G_{BS}}} {h_{2b}}{b_k} + {n_1},
\end{split}
\end{equation}
where ${d_{1b}}$ and ${d_{2b}}$ are the transmission distance from U1 to BS and that from U2 to BS, respectively, ${h_{1b}}$ and ${h_{2b}}$ are fading coefficients of the channel from U1 to BS and that from U2 to BS, respectively, and ${n_{1}}$ is a circularly symmetric complex Gaussian (CSCG) random variable with zero mean and noise variance ${N_0}{B_c}$. Then, during the second half of time slot $k$, U1 and U2 transmit $- b_k^*$ and $a_k^*$, respectively, where $*$ denotes the conjugate operation. Thus, the signal received at BS is expressed as
\begin{equation}\label{equa6}
\begin{split}
{y_2} = & - \sqrt {{P_{1,c}}{(\dfrac{{{\lambda _c}}}{{4\pi {d_{1b}}}})^2}{G_{U1}}{G_{BS}}} {h_{1b}}b_k^* \\
&+ \sqrt {{P_{2,c}}{(\dfrac{{{\lambda _c}}}{{4\pi {d_{2b}}}})^2}{G_{U2}}{G_{BS}}} {h_{2b}}a_k^* + {n_2},
\end{split}
\end{equation}
where ${n_2}$ is a CSCG random variable with zero mean and noise variance ${N_0}{B_c}$. From (5) and (6), BS can decode ${a_k}$ and ${b_k}$ separately by using Alamouti decoding algorithm. To be specific, after Alamouti decoding, the same received SNR is achieved at BS in decoding both ${a_k}$ and ${b_k}$, which is given by
\begin{equation}\label{equa7}
\begin{split}
&\gamma _{1b}^{NC}(\theta  = 1) = \gamma _{2b}^{NC}(\theta  = 1)\\
& = \dfrac{{{P_{1,c}}}}{{{N_0}{B_c}}}{\left(\dfrac{{{\lambda _c}}}{{4\pi {d_{1b}}}}\right)^2}{G_{U1}}{G_{BS}}|{h_{1b}}{|^2}\\
& \quad + \dfrac{{{P_{2,c}}}}{{{N_0}{B_c}}}{\left(\dfrac{{{\lambda _c}}}{{4\pi {d_{2b}}}}\right)^2}{G_{U2}}{G_{BS}}|{h_{2b}}{|^2}.
\end{split}
\end{equation}
In the case of $\theta  = 2$, i.e., either U1 or U2 (or both) fails to decode the short-range transmissions, we allow U1 and U2 to transmit their signals ${a_k}$ and ${b_k}$ to BS separately over a cellular spectrum band. For example, during the first half of time slot $k$, U1 transmits ${a_k}$ to BS with power ${P_{1,c}}$ and then U2 transmits ${b_k}$ to BS with power ${P_{2,c}}$ during the second half of slot $k$. Therefore, in the case of $\theta  = 2$, the received SNRs at BS in decoding ${a_k}$ and ${b_k}$ (from U1 and U2) are, respectively, given by
\begin{equation}\label{equa8}
\gamma _{1b}^{NC}(\theta  = 2) = \dfrac{{{P_{1,c}}}}{{{N_0}{B_c}}}{\left(\dfrac{{{\lambda _c}}}{{4\pi {d_{1b}}}}\right)^2}{G_{U1}}{G_{BS}}|{h_{1b}}{|^2},
\end{equation}
and
\begin{equation}\label{equa9}
\gamma _{2b}^{NC}(\theta  = 2) = \dfrac{{{P_{2,c}}}}{{{N_0}{B_c}}}{\left(\dfrac{{{\lambda _c}}}{{4\pi {d_{2b}}}}\right)^2}{G_{U2}}{G_{BS}}|{h_{2b}}{|^2}.
\end{equation}
This completes the signal model of the inter-network cooperation scheme.

\subsection{Conventional Intra-Network Cooperation}
For the purpose of comparison, this subsection presents the conventional intra-network cooperation scheme [4], [5]. Similarly, we consider U1 and U2 that transmit ${a_k}$ and ${b_k}$ to BS, respectively. In the conventional intra-network cooperation scheme [5], U1 and U2 first exchange their signals (i.e., ${a_k}$ and ${b_k}$) between each other over cellular spectrum bands, which is different from the inter-network cooperation case where the information exchanging operates in a short-range communication network over ISM bands. During the information exchange process, U1 and U2 attempt to decode each other's signals. If both U1 and U2 successfully decode, the Alamouti space-time coding is used in transmitting ${a_k}$ and ${b_k}$ from U1 and U2 to BS over cellular bands. Otherwise, U1 and U2 transmit ${a_k}$ and ${b_k}$ to BS, separately. Note that the conventional scheme requires two orthogonal phases to complete each packet transmission, which causes the loss of one-half of cellular spectrum utilization. Thus, the conventional scheme needs to transmit at twice of the data rate of the inter-network cooperation scheme in order to send the same amount of information. In other words, U1 and U2 should transmit ${a_k}$ and ${b_k}$ at data rates $2R_1$ and $2R_2$, respectively, for a fair comparison. One can observe that the signal model of the conventional intra-network cooperation is almost the same as that of the inter-network cooperation, except that $2R_1$ and $2R_2$ are considered as the data rates of U1 and U2 in the conventional scheme and, moreover, the information exchange between U1 and U2 in the conventional scheme operates over cellular bands instead of ISM bands.

\section{Energy Efficiency Analysis with Target Outage and Rate Requirements}
In this section, we analyze the energy efficiency of the traditional non-cooperation, conventional intra-network cooperation, and proposed inter-network cooperation schemes by assuming that different users (i.e., U1 and U2) have the same target outage probability and data rate requirements.
\subsection{Traditional Non-cooperation}
Without loss of generality, let $\overline {{\rm{Pout}}} $ and $\overline R $ (in ${\textrm{bits/s}}$) denote the common target outage probability and data rate, respectively, for both users. Using (1) and assuming the optimal Gaussian codebook, the maximum achievable rate from U1 to BS is given by
\begin{equation}\label{equa10}
C_{1b}^T = {B_c}{\log _2}[1 + \dfrac{{{P_1}}}{{{N_0}{B_c}}}{(\dfrac{{{\lambda _c}}}{{4\pi {d_{1b}}}})^2}{G_{U1}}{G_{BS}}|{h_{1b}}{|^2}].
\end{equation}
As we know, an outage event occurs when the channel capacity falls below the data rate. Note that the random variable $|{h_{1b}}{|^2}$ follows an exponential distribution with mean $\sigma _{1b}^2$. Thus, we can compute the outage probability for U1's transmission as
\begin{equation}\label{equa11}
\begin{split}
{\rm{Pout}}_1^T &= \Pr (C_{1b}^T < {R_1}) \\
&= 1 - \exp [ - \dfrac{{16{\pi ^2}{N_0}{B_c}d_{1b}^2({2^{{R_1}/{B_c}}} - 1)}}{{{P_1}\sigma _{1b}^2{G_{U1}}{G_{BS}}\lambda _c^2}}].
\end{split}
\end{equation}
Given the target outage and rate requirements, i.e., ${\rm{Pout}}_1^T = \overline {{\rm{Pout}}} $ and ${R_1} = \overline R $, we can easily compute the power consumption of U1 from (11) as ${P_1} =  - \frac{{16{\pi ^2}{N_0}{B_c}d_{1b}^2({2^{\overline R /{B_c}}} - 1)}}{{\sigma _{1b}^2{G_{U1}}{G_{BS}}\lambda _c^2\ln (1 - \overline {{\rm{Pout}}} )}}$. Similarly, we can obtain the power consumption of U2 as ${P_2} =  - \frac{{16{\pi ^2}{N_0}{B_c}d_{2b}^2({2^{\overline R /{B_c}}} - 1)}}{{\sigma _{2b}^2{G_{U2}}{G_{BS}}\lambda _c^2\ln (1 - \overline {{\rm{Pout}}} )}}$. Thus, considering the traditional non-cooperation, the total power consumption of both U1 and U2 is obtained as ${P_T} = {P_1} + {P_2}$, based on which a so-called energy efficiency in Bits-per-Joule is defined as
\begin{equation}\label{equa12}
{\eta_T} = \dfrac{ \overline R}{P_T},
\end{equation}
where $\overline R$ in ${\textrm{bits/s}}$ is the target data rate requirement. Notice that the energy efficiency ${\eta_T}$ is used to quantify the number of bits transmitted per unit energy.

\subsection{Proposed Inter-Network Cooperation}
In this subsection, we present an energy efficiency analysis of the proposed inter-network cooperation. Notice that all random variables $|{h_{12}}{|^2}$, $|{h_{21}}{|^2}$, $|{h_{1b}}{|^2}$ and $|{h_{2b}}{|^2}$ follow independent exponential distributions with means $\sigma _{12}^2$, $\sigma _{21}^2$, $\sigma _{1b}^2$ and $\sigma _{2b}^2$, respectively. From (3), we can obtain the outage probability of the short-range transmission from U1 to U2 as
\begin{equation}\label{equa13}
\begin{split}
{\rm{Pou}}{{\rm{t}}_{12}} &= \Pr (\gamma _{12}^{NC} < {2^{{R_1}/{B_s}}} - 1) \\
&= 1 - \exp [ - \dfrac{{16{\pi ^2}{N_0}{B_s}d_{12}^2({2^{{R_1}/{B_s}}} - 1)}}{{{P_{1,s}}\sigma _{12}^2{G_{U1}}{G_{U2}}\lambda _s^2}}].
\end{split}
\end{equation}
Assuming ${\rm{Pou}}{{\rm{t}}_{12}} = \overline {{\rm{Pout}}} $ and ${R_1} = \overline R $, the power consumption of U1 for short-range communication is given by ${P_{1,s}} =  - \frac{{16{\pi ^2}{N_0}{B_s}d_{12}^2({2^{\overline R /{B_s}}} - 1)}}{{\sigma _{12}^2{G_{U1}}{G_{U2}}\lambda _s^2\ln (1 - \overline {{\rm{Pout}}} )}}$. From (4), we similarly obtain the power consumption of U2 for short-range communication as ${P_{2,s}} =  - \frac{{16{\pi ^2}{N_0}{B_s}d_{21}^2({2^{\overline R /{B_s}}} - 1)}}{{\sigma _{21}^2{G_{U1}}{G_{U2}}\lambda _s^2\ln (1 - \overline {{\rm{Pout}}} )}}$.
In addition, according to (7) and (8), we obtain the outage probability of U1's transmission with the inter-network cooperation as
\begin{equation}\label{equa14}
\begin{split}
{\rm{Pout}}_1^{NC} &= \Pr (\theta  = 1)\Pr [\gamma _{1b}^{NC}(\theta  = 1) < {2^{{R_1}/{B_c}}} - 1] \\
&+ \Pr (\theta  = 2)\Pr [\gamma _{1b}^{NC}(\theta  = 2) < {2^{{R_1}/{B_c}}} - 1].
\end{split}
\end{equation}
As discussed in section II-B, case $\theta  = 1$ implies that both U1 and U2 succeed in decoding each other's signals through short-range communications, and $\theta  = 2$ means that either U1 or U2 (or both) fails to decode in the short-range transmissions. Specifically, we can describe the two events $\theta  = 1$ and $\theta  = 2$ as follows
\begin{equation}\label{equa15}
\begin{split}
&\theta = 1:{\textrm{ }}{B_s}{\log _2}(1 + \gamma _{12}^{NC}) > {R_1}{\textrm{ and }}{B_s}{\log _2}(1 + \gamma _{21}^{NC}) > {R_2}\\
&\theta = 2:{\textrm{ }}{B_s}{\log _2}(1 + \gamma _{12}^{NC}) < {R_1}{\textrm{  or  }}{B_s}{\log _2}(1 + \gamma _{21}^{NC}) < {R_2}.
\end{split}
\end{equation}
Assuming $\overline {{\rm{Pout}}} $ for short-range communication between U1 and U2, we have $\Pr (\theta  = 1)= {(1 - \overline {{\rm{Pout}}} )^2}$
and $\Pr (\theta  = 2) = 1 - {(1 - \overline {{\rm{Pout}}} )^2}$. Notice that random variables $|{h_{1b}}{|^2}$ and $|{h_{2b}}{|^2}$ are independent and both follow exponential distributions with respective mean $\sigma _{1b}^2$ and $\sigma _{2b}^2$. Similarly to (55) in [12], we can obtain a closed-form expression of $\Pr [\gamma _{1b}^{NC}(\theta  = 1) < {2^{{R_1}/{B_c}}} - 1]$ by using (7). Besides, using (8), we can easily express $\Pr [\gamma _{1b}^{NC}(\theta  = 2) < {2^{{R_1}/{B_c}}} - 1]$ in (14) as
\begin{equation}\label{equa16}
\begin{split}
&\Pr [\gamma _{1b}^{NC}(\theta  = 2) < {2^{{R_1}/{B_c}}} - 1] \\
&= 1 - \exp [ - \dfrac{{16{\pi ^2}{N_0}{B_c}d_{1b}^2({2^{{R_1}/{B_c}}} - 1)}}{{{P_{1,c}}\sigma _{1b}^2{G_{U1}}{G_{BS}}\lambda _c^2}}].
\end{split}
\end{equation}
Similarly to (14), we can obtain from (7) and (9) the outage probability of U2's transmissions as
\begin{equation}\label{equa17}
\begin{split}
{\rm{Pout}}_2^{NC}& = \Pr (\theta  = 1)\Pr [\gamma _{2b}^{NC}(\theta  = 1) < {2^{{R_2}/{B_c}}} - 1] \\
&+ \Pr (\theta  = 2)\Pr [\gamma _{2b}^{NC}(\theta  = 2) < {2^{{R_2}/{B_c}}} - 1],
\end{split}
\end{equation}
where $\Pr (\theta  = 1)= {(1 - \overline {{\rm{Pout}}} )^2}$ and $\Pr (\theta  = 2) = 1 - {(1 - \overline {{\rm{Pout}}} )^2}$. Moreover, similarly to (55) of [12] and (16), we easily determine $\Pr [\gamma _{2b}^{NC}(\theta  = 1) < {2^{{R_2}/{B_c}}} - 1]$ and $\Pr [\gamma _{2b}^{NC}(\theta  = 2) < {2^{{R_2}/{B_c}}} - 1]$ in closed-form. Considering ${\rm{Pout}}_1^{NC} = {\rm{Pout}}_2^{NC} = \overline {{\rm{Pout}}} $ and ${R_1} = {R_2} = \overline R $ and using (14) and (17), we have
\begin{equation}\label{equa18}
{P_{2,c}} = \frac{{\sigma _{1b}^2{G_{U1}}d_{2b}^2}}{{\sigma _{2b}^2{G_{U2}}d_{1b}^2}}{P_{1,c}}.
\end{equation}
Therefore, letting ${\rm{Pout}}_1^{NC} = \overline {{\rm{Pout}}} $ and ${R_1} = \overline R $ and substituting (18) into (14), we obtain
\begin{equation}\label{equa19}
\begin{split}
{P_{1,c}}& = \frac{{16{\pi ^2}{N_0}{B_c}d_{1b}^2({2^{\overline R /{B_c}}} - 1)}}{{\sigma _{1b}^2\lambda _c^2{G_{U1}}{G_{BS}}}}\\
&\times{\left[ { - \frac{1}{{{{(1 - \overline {{\rm{Pout}}} )}^2}}} - W\left( {\frac{{\exp [ - {{(1 - \overline {{\rm{Pout}}} )}^{ - 2}}]}}{{\overline {{\rm{Pout}}}  - 1}}} \right)} \right]^{ - 1}},
\end{split}
\end{equation}
and
\begin{equation}\label{equa20}
\begin{split}
{P_{2,c}} &=\frac{{16{\pi ^2}{N_0}{B_c}d_{2b}^2({2^{\overline R /{B_c}}} - 1)}}{{\sigma _{2b}^2\lambda _c^2{G_{U2}}{G_{BS}}}}\\
&\times{\left[ { - \frac{1}{{{{(1 - \overline {{\rm{Pout}}} )}^2}}} - W\left( {\frac{{\exp [ - {{(1 - \overline {{\rm{Pout}}} )}^{ - 2}}]}}{{\overline {{\rm{Pout}}}  - 1}}} \right)} \right]^{ - 1}},
\end{split}
\end{equation}
where $W( \cdot )$ is the lambert function. Notice that in case of $\theta  = 1$, the Alamouti space-time coding is employed, resulting in that a total power of $2({P_{1,c}} + {P_{2,c}})$ is consumed by U1 and U2 in transmitting to BS. In case of $\theta  = 2$, U1 and U2 consume a total power of $({P_{1,c}} + {P_{2,c}})$ in transmitting to BS. Therefore, considering both the short-range communication and cellular transmissions, the total power consumption by the inter-network cooperation is given by
\begin{equation}\label{equa21}
\begin{split}
{P_{NC}} &= {P_{1,s}} + {P_{2,s}} + 2\Pr (\theta  = 1)({P_{1,c}} + {P_{2,c}}) \\
&\quad+ \Pr (\theta  = 2)({P_{1,c}} + {P_{2,c}})\\
&= {P_{1,s}} + {P_{2,s}} + [1+(1-\overline {{\rm{Pout}}})^2]({P_{1,c}} + {P_{2,c}}), \\
\end{split}
\end{equation}
where ${P_{1,c}}$ and ${P_{2,c}}$ are given by (19) and (20), respectively. Based on (21), the energy efficiency of proposed network cooperation is given by
\begin{equation}\label{equa22}
{\eta_{NC}} = \dfrac{\overline R}{P_{NC}},
\end{equation}
where ${P_{NC}}$ is given in (21).

\subsection{Conventional Intra-Network Cooperation}
As discussed in Section II-D, the intra-network cooperation differs from the proposed inter-network cooperation in two main aspects. First, in the intra-network cooperation case, U1 and U2 transmit at rates $2R_1$ and $2R_2$, respectively, to make the same effective transmission rates as the inter-network cooperation case. This means that in characterizing the energy efficiency of intra-network cooperation, we need to replace $R$ in the energy efficiency expressions of inter-network cooperation, i.e., (19) and (20), with $2R$. Secondly, in the intra-network cooperation, the information exchange between U1 and U2 operates over cellular bands instead of ISM bands. Therefore, the energy consumed in the information exchange phase of intra-network cooperation can be obtained from (13) by replacing $\lambda_s$ and $B_s$ with $\lambda_c$ and $B_c$, respectively. In this way, the energy efficiency of conventional intra-network cooperation can be readily determined.

\section{Numerical Results}
\begin{table}
  \centering
  \caption{System Parameters Used in Numerical Evaluation.}
  {\includegraphics[scale=0.7]{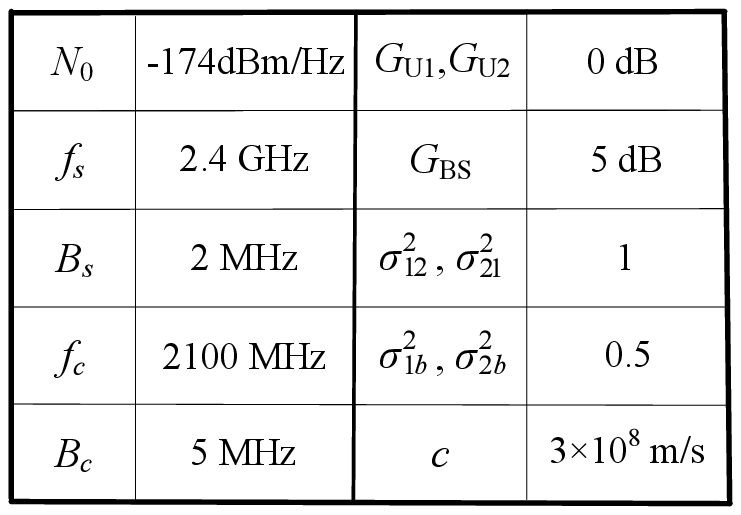}\label{Tab1}}
\end{table}

\begin{figure}
  \centering
  {\includegraphics[scale=0.6]{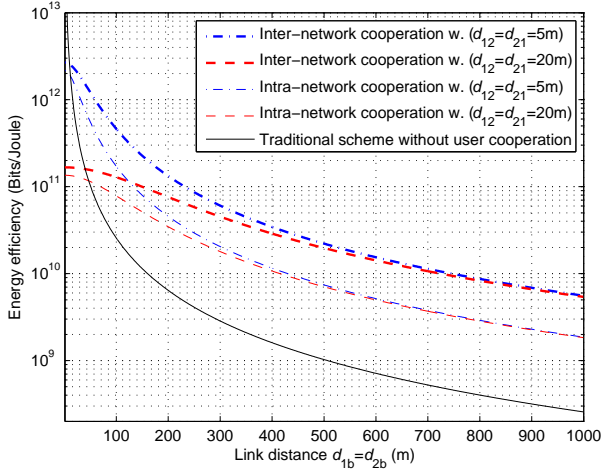}\\
  \caption{Energy efficiency versus link distance from U1/U2 to BS (${d_{1b}} = {d_{2b}}$) of various transmission schemes with target outage probability $\overline {{\rm{Pout}}}  = {10^{ - 3}}$ and data rate $\overline R  = 5M{\textrm{bits/s}}$.}\label{Fig4}}
\end{figure}

In this section, we present numerical results on energy efficiency of various transmission schemes given the target outage probability and data rate requirements. Table I summarizes the system parameters used in the numerical evaluation. Fig. 4 shows the energy efficiency comparison among the traditional non-cooperation, intra-network cooperation, and inter-network cooperation by plotting (12) and (22) as a function of link distance ${d_{1b}} = {d_{2b}}$ in meters ($m$). One can observe from Fig. 4 that for both cases of $d_{12}=d_{21}=5m$ and $d_{12}=d_{21}=20m$, the intra- and inter-network cooperation schemes achieve higher energy efficiency than the traditional non-cooperation. This means that given a certain amount of bits to be transmitted, the intra- and inter-network cooperation schemes require less energy as compared with the non-cooperation scheme, showing the energy saving benefit of exploiting user cooperation. Fig. 4 also shows that for both cases of $d_{12}=d_{21}=5m$ and $d_{12}=d_{21}=20m$, the energy efficiency of proposed inter-network cooperation is higher than that of conventional intra-network cooperation.

\begin{figure}
  \centering
  {\includegraphics[scale=0.6]{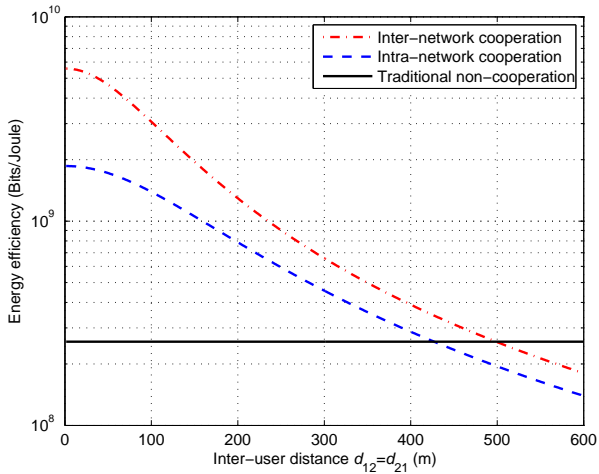}\\
  \caption{Energy efficiency versus inter-user distance between U1 and U2 of various transmission schemes with target outage probability $\overline {{\rm{Pout}}}  = {10^{ - 3}}$, data rate $\overline R  = 5M{\textrm{bits/s}}$, and ${d_{1d}}={d_{2d}}=1000m$.}\label{Fig5}}
\end{figure}
Fig. 5 depicts the energy efficiency versus inter-user distance between U1 and U2 of the traditional non-cooperation, intra-network cooperation, and inter-network cooperation schemes. As shown in Fig. 5, the energy efficiency of traditional non-cooperation is constant in this case, which is due to the fact that ${\eta_T}$ given in (12) is independent of the inter-user distance. One can see from Fig. 5 that when the inter-user distance is small, both the intra- and inter-network cooperation significantly outperform the traditional non-cooperation in terms of energy efficiency. However, as the inter-user distance increases beyond a certain value, the intra- and inter-network cooperation perform worse than the traditional non-cooperation, showing that user cooperation is not energy efficient when U1 and U2 are far away from each other. In addition, it is shown from Fig. 5 that the energy efficiency of proposed inter-network cooperation is always better than that of intra-network cooperation.

\section{Conclusion}
In this paper, we studied the multiple network access interfaces assisted user cooperation, termed inter-network cooperation, to improve the energy efficiency of cellular uplink transmissions. We derived closed-form expressions of the energy efficiency of traditional non-cooperation, conventional intra-network cooperation, and proposed inter-network cooperation with target outage probability and data rate requirements. Numerical results showed that as the cooperating users move towards to each other, the proposed inter-network cooperation substantially improves the energy efficiency over traditional non-cooperation and intra-network cooperation, showing the advantage of inter-network cooperation.

\ifCLASSOPTIONcaptionsoff
  \newpage
\fi

\end{document}